\documentclass[trackchanges,twocolumn]{aastex701}

\usepackage{amsmath}
\usepackage{tabularx}
\usepackage{makecell}
\usepackage{adjustbox}
\usepackage{subcaption}
\definecolor{orcidlogocol}{HTML}{A6CE39}
\newcommand{\orcidicon}[1]{\href{https://orcid.org/#1}{\textcolor{orcidlogocol}{\aiOrcid}}}

\begin{document}

\title{Dynamics and Radiative Signatures of Accretion Flows onto a Kerr-like Wormhole}

\correspondingauthor{Jing-ze Xia, Hong-Xuan Jiang, \\ Yosuke Mizuno}
\email{jackxia@sjtu.edu.cn, hongxuan\_jiang@sjtu.edu.cn, \\mizuno@sjtu.edu.cn}

\author[0009-0004-8669-2411]{Jing-Ze Xia} 
\affiliation{Tsung-Dao Lee Institute, Shanghai Jiao Tong University,
1 Lisuo Road, Shanghai 201210, PR China}
\email{jackxia@sjtu.edu.cn}

\author[0000-0003-0292-2773]{Hong-Xuan Jiang}
\affiliation{Tsung-Dao Lee Institute, Shanghai Jiao Tong University,
1 Lisuo Road, Shanghai 201210, PR China}
\email{hongxuan_jiang@sjtu.edu.cn}

\author[0000-0002-2714-9257]{Cheng Liu}
\affiliation{Tsung-Dao Lee Institute, Shanghai Jiao Tong University,
1 Lisuo Road, Shanghai 201210, PR China}
\email{liuc09@sjtu.edu.cn}

\author[0000-0002-8131-6730]{Yosuke Mizuno}
\affiliation{Tsung-Dao Lee Institute, Shanghai Jiao Tong University,
1 Lisuo Road, Shanghai 201210, PR China}
\affiliation{School of Physics and Astronomy, Shanghai Jiao Tong University,
800 Dongchuan Road, Shanghai 200240, PR China}
\affiliation{Key Laboratory for Particle Physics, Astrophysics and Cosmology (MOE),
Shanghai Key Laboratory for Particle Physics and Cosmology,
Shanghai Jiao Tong University, 800 Dongchuan Road, Shanghai 200240, PR China}
\email{mizuno@sjtu.edu.cn}

\begin{abstract}
Wormholes are a hypothetical object that connects disparate points in spacetime. It is a theoretically well-motivated black hole alternative and offers a potential observationally testable arena for probing strong-field gravity with horizon-scale images.
We perform general relativistic magnetohydrodynamic (GRMHD) simulations and general relativistic radiative transfer (GRRT) calculations of accretion flows onto a Kerr-like wormhole. Adopting a Kerr black-bounce metric with a fixed throat parameter $\ell = 2.5\,\rm M$, we explore the effects of spin using both two- and three-dimensional simulations. The accretion flow is initialized as a magnetized geometrically thick torus near one mouth of the wormhole, while the opposite mouth is initially gas-free. We find that the spin parameter influences the dynamical properties on both sides of the wormhole through the frame-dragging effects.
Based on the GRMHD results, we compute ray-traced images at $230\,\mathrm{GHz}$ using \texttt{RAPTOR}, and analyze the horizon-scale image structure through higher-order photon trajectories. 
Our GRRT calculations show that emissions originating from the immediate vicinity of the throat can dominate, in contrast to the case of a Kerr black hole.
It provides the variable component of the signal and imprints a clear quasi-periodic modulation in the light curves. 
These properties would be useful to either confirm or rule out such exotic compact objects through horizon-scale observations.
\end{abstract}


\section{Introduction}
Horizon-scale imaging of supermassive compact objects has opened a direct observational window onto the strong-field regime of gravity. In particular, the Event Horizon Telescope (EHT) \citep{2019ApJ...875L...5E,2022ApJ...930L..12E} has resolved ring-like structures around M~87 and Sgr~A$^{*}$, enabling quantitative tests of spacetime geometry through shadow size, asymmetry, and substructure at sub-horizon scales. These developments have intensified interest in ``black-hole mimickers''---horizonless compact objects that can reproduce Kerr-like appearances while potentially leaving detectable imprints in imaging and variability observables \citep[e.g.,][]{2021PhRvD.103j4050K,2025arXiv251120756S,2020MNRAS.497..521O,2025ApJ...978...44D,2019ApJ...875L...6E,2021A&A...646A..37V,2018PhRvD..98b4044S,Cheng:2026wyk}.

Among the proposed alternatives, traversable wormholes provide a conceptually distinct class of horizonless spacetimes \citep{2019JCAP...02..042S}. The black-bounce framework offers a simple and continuous interpolation between black holes and traversable wormholes, and its rotating generalizations \citep{Mazza:2021rgq} yield Kerr-like wormhole geometries that can mimic a Kerr black hole at large radii while modifying the central photon dynamics through a finite throat scale \citep{2022PhRvD.105h4036T,2023PhRvD.108h4054I}. Such Kerr-like wormholes, therefore, form a useful laboratory for assessing how the horizonless structure may manifest itself in horizon-scale observations.

While the construction of such traversable wormhole spacetimes typically necessitates the violation of standard energy conditions through the introduction of exotic matter at the throat, our approach here is primarily phenomenological. By adopting the black-bounce framework as a well-defined mathematical laboratory, we treat the throat scale $l$ as a free parameter to explore the observational limits of horizonless objects. 
A central geometric diagnostic for imaging is the critical curve in the observer's sky, which is determined by the separatrix of photon trajectories \citep[e.g.,][]{1973blho.conf..215B}. For Kerr-like wormholes, the presence of a throat can qualitatively alter the critical curve: besides the usual Kerr-like branch associated with unstable spherical photon orbits, an additional throat-controlled branch may appear when the throat scale competes with the location of the innermost unstable photon orbits \citep{2021PhRvD.103j4050K}. This throat effect modifies the shadow boundary and, more importantly, the accumulation locus of higher-order photon trajectories that build up the photon-ring substructure \citep{2021A&A...646A..37V}.

Connecting these geometric predictions to realistic observables requires a self-consistent treatment of plasma dynamics near the central compact object and radiation \citep[e.g.,][]{2019ApJ...875L...5E,2019ApJS..243...26P,2022ApJ...930L..12E,Wong:2022rqr}. Recent GRMHD studies of traversable wormholes based on the non-rotating Simpson-Visser metric \citep{Combi:2024ehi} have shown that accretion through a throat can naturally lead to matter accumulation and enhanced dissipation in the central region, potentially producing bright in the vicinity of the throat. Complementary work on disk emission around Kerr-mimicking wormholes further highlights that even when time-averaged spectra or temperatures resemble Kerr predictions, horizonless structure can still imprint signatures that become more apparent in carefully chosen observables \citep{2025arXiv251207466K}.

Motivated by these considerations, in this paper, we perform a set of two- and three-dimensional GRMHD simulations of magnetized accretion flows in a Kerr-like wormhole spacetime and carry out general relativistic radiative transfer (GRRT) calculations based on the dynamical flow solutions. 
In this study, we fix the throat parameter and vary the spin. We examine how rotation influences magnetization and outflows properties \citep{Urtubey:2024pkm, 2025EPJC...85.1178E}.
We perform a direct comparison between analytically derived critical curves and higher-order photon rings identified in GRRT images. We further explore time-dependent radiative signatures and discuss their potential relevance for future horizon-scale imaging and variability constraints.

This paper is organized as follows. In Section~2, we summarize the Kerr black-bounce metric and our numerical setup for GRMHD simulations and GRRT calculations. Section~3 presents the GRMHD results for the accretion and outflow structures. In Section~4, we analyze the ray-traced images and light curves. We summarize our findings and discuss the observational implications and future directions in Section~5.

\section{simulation setup}
\subsection{Metric}

We first briefly review the Kerr black-bounce metric ~\citep{Mazza:2021rgq}:
\begin{equation}
\begin{aligned}
ds^2 ={}& -\left(1-\frac{2M\sqrt{   r^{\,2}+\ell^2}}{\Sigma}\right)dt^2
+ \frac{\Sigma}{\Delta}\,d   r^{\,2}
+ \Sigma\, d\theta^2  \\
&+ \frac{A\sin^2\theta}{\Sigma}\,d\phi^2
- \frac{4Ma\sqrt{   r^{\,2}+\ell^2}\sin^2\theta}{\Sigma}\,dt\,d\phi .
\end{aligned}
\label{eq:kbounce_metric}
\end{equation}
Here, \(M\), \(a\), and \(\ell\) denote the mass, spin parameter, and the regularization parameter of the spacetime, respectively. In the limit of a vanishing (positive) regularization parameter, $\ell \to 0$, the Kerr--black-bounce metric smoothly reduces to the (singular) Kerr solution. For $\ell \neq 0$, spacetime is everywhere regular and features a wormhole throat at $r=0$.
The metric functions are defined as
\begin{equation}
\begin{aligned}
\Sigma &=    r^{\,2} + \ell^2 + a^2\cos^2\theta, \\
\Delta &=    r^{\,2} + \ell^2 + a^2 - 2M\sqrt{   r^{\,2}+\ell^2}, \\
A      &= (   r^{\,2} + \ell^2 + a^2)^2 - \Delta a^2\sin^2\theta .
\end{aligned}
\label{eq:kbounce_functions}
\end{equation}

This spacetime represents a rotating generalization of the static and spherically symmetric black-bounce metric originally proposed by Simpson and Visser~\citep{Simpson:2018tsi,Lobo:2020ffi}. In the limit $\ell \to 0$, the spacetime returns smoothly to the Kerr solution. In our simulations, we fix the regularizing parameter to $\ell = 2.5$ and vary the spin parameter over the values $a = 0$, $0.3$, and $0.9$.

\subsection{Numerical setup}
\subsubsection{GRMHD setup}

Our numerical setup closely follows that adopted in~\cite{Combi:2024ehi}. The GRMHD equations are evolved using the GPU-based code \texttt{KHARMA}~\citep{Prather:2024hsu}, descended from the original \texttt{HARM} scheme, which solves the equations of ideal magnetohydrodynamics in a general relativistic framework \citep{Gammie:2003rj}. This framework enables simulations of accretion flows in arbitrary spacetimes and coordinate systems.

We construct a spherical computational grid that is symmetric about the wormhole throat, with an increase in radial resolution in its vicinity. The uniform numerical coordinates $(x_1, x_2, x_3)$ are mapped to the physical coordinates by
$r(x_1) = c_r \sinh(x_1 / c_r)$,
where we fix $c_r = 1 - \log 2 / \log R_{\rm out}$ and set the outer radial boundary at $R_{\rm out} = 1000$~\citep{Combi:2024ehi}. The angular coordinates are chosen as $\theta(x_2) = x_2$ and $\phi(x_3) = x_3$. The radial domain extends over $r \in [-1000M,\,1000M]$, where the outflow boundary conditions are imposed at both radial boundaries.  We cover the entire $(\theta,\phi)$ sphere. Transmissive boundary conditions are applied at the angular boundaries, following~\cite{Combi:2024ehi}.
We assume that the wormhole accretes gas through mouth A from a hot, geometrically thick torus, and the opposite mouth (mouth B) is initially taken to be gas-free. We adopt an ideal-gas equation of state,
$u = p/(\Gamma - 1)$, where $u$ is the internal energy density and $p$ is the gas pressure. 
In this work, we adopt an adiabatic index of $\gamma = 4/3$ primarily to
  facilitate direct comparison with previous studies, particularly
  \cite{Combi:2024ehi}. This choice also reflects our emphasis on the
  relativistic nature of the near-horizon flow, as discussed by
  \citet{2025ApJ...980..193G}, that electrons in low-luminosity black hole accretion flows can
  become mildly relativistic near the horizon even when the ions remain non-relativistic.

We initialize the simulation with a generalized Fishbone--Moncrief (FM) torus placed near mouth A~\citep{Uniyal:2024sdv}. The inner edge of the torus is located at $r = 25M$, and the maximum density of the disk occurs at $r = 35M$. In addition, a weak poloidal magnetic loop is initialized within the torus, its strength being specified by a minimum plasma beta parameter of $\beta = 150$. To trigger magnetorotational instability (MRI), we add a random perturbation to the internal energy of the initial torus
$u \rightarrow u + \delta u$,
$|\delta u|/u \le u_{\rm jitter}$,
following the prescription of~\cite{Wong:2022rqr}. In this work, we adopt $u_{\rm jitter} = 0.1$. On the opposite side (mouth B), all MHD variables are initially set to their atmosphere values. 

The resolutions and spin parameters used in our simulations are summarized in Table~\ref{tab:table1}. Compared to \cite{Combi:2024ehi}, they used a stationary metric with $a=0$ with $\ell=2.1$ and $3$. In our simulations, we adopt $\ell = 2.5$ as an intermediate representative value for the throat parameter with three different spin values. A discussion of the MRI quality factors and their implications for numerical resolution are provided in the Appendix~\ref{appenD}.

\subsubsection{GRRT setup}

Using the public ray-tracing code GRRT \texttt{RAPTOR}, we compute images based on our GRMHD simulations of the wormhole spacetime \citep{2018A&A...613A...2B}. The source is assumed to be Sgr~A$^{*}$, with a mass of $M = 4.14 \times 10^{6}\,M_\odot$ and located at a distance of $8.127\,\mathrm{kpc}$. Here, we consider synchrotron radiation with a thermal electron distribution function as the emission mechanism. For simplicity, we adopt parameterized R-$\beta$ model with $R_{\rm low} =1,\, R_{\rm high} = 10$ \citep{2019ApJ...875L...5E,2022ApJ...930L..12E,2023MNRAS.522.2307J,2024A&A...688A..82J} to obtain electron-to-ion temperature ratio from GRMHD simulations. We compute ray-traced images on an image plane with a resolution of $1000 \times 1000$ pixels with a field of view $\pm30\,r_g$ at an observing frequency of $230\,\mathrm{GHz}$, matching the EHT observing band.

\begin{table}[b]
\caption{\label{tab:table1}%
The settings of different runs.
}
\begin{ruledtabular}
\begin{tabular}{lcc}
\textrm{Models}&
\textrm{Resolution($n_r\times n_\theta \times n_\phi)$}&
\textrm{Spin}\\
\colrule
\tt{M2Da03} & $4096\times 1024\times 1$ & $a=0.3 $\\
\tt{M2Da09}& $4096\times 1024\times 1$ & $a=0.9 $ \\
\tt{M3Da00} & $512\times 120\times 96 $ & $a=0.0 $ \\
\tt{M3Da03} & $512\times 120\times 96 $ & $a=0.3 $ \\
\tt{M3Da09} & $512\times 120\times 96 $ & $a=0.9 $ \\
\end{tabular}
\end{ruledtabular}
\end{table}

\begin{figure*}
\includegraphics[width=\textwidth]{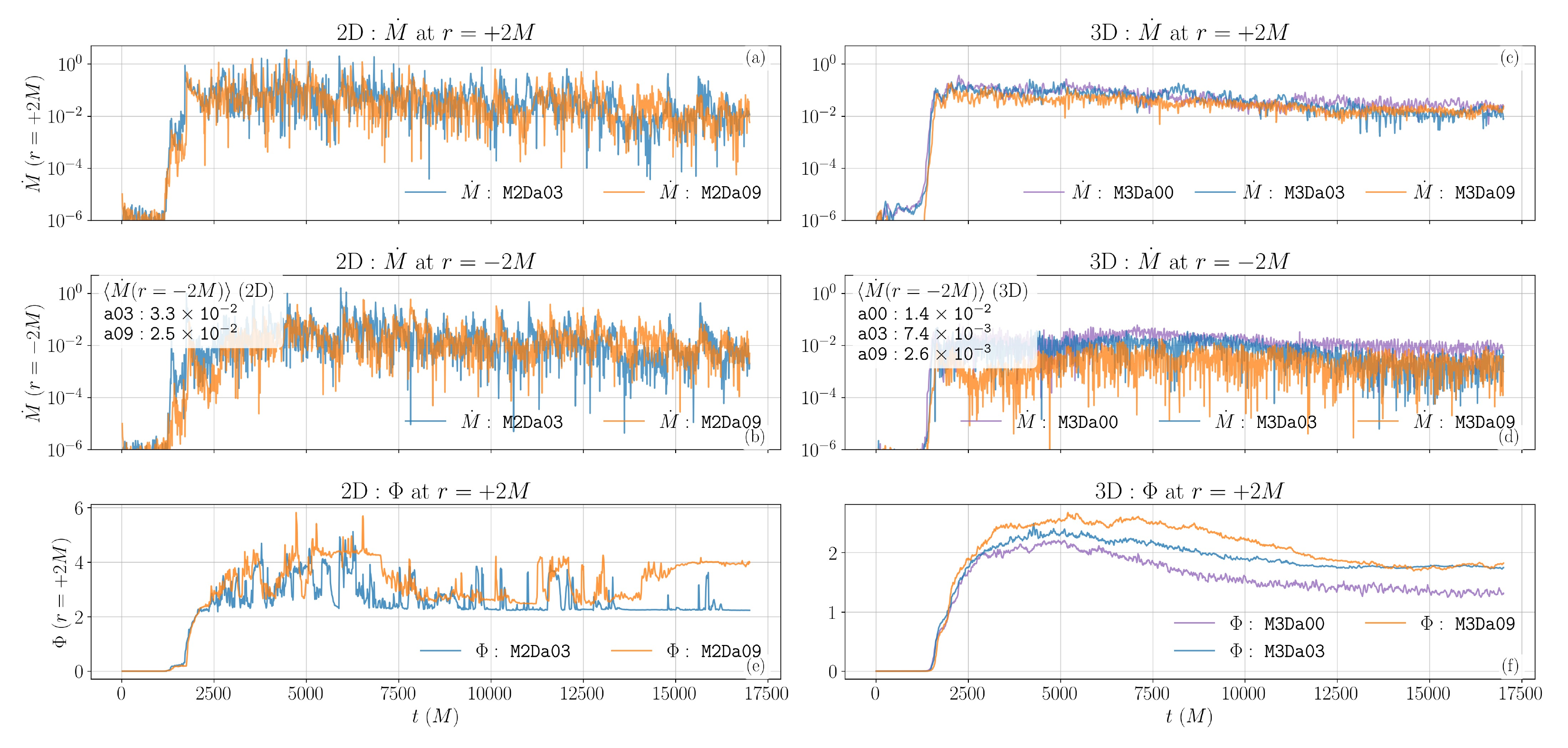}
\caption{The mass flux rate at different mouth (top: $r=2\,\rm M$ and middle: $r=-2\,\rm M$) from 2D GRMHD simulations, $\tt{M2Da03}$ and $\tt{M2Da09}$ (left), and 3D cases $\tt{M3Da00}$, $\tt{M3Da03}$, and $\tt{M3Da09}$ (right), respectively. We note that mass flux is inflow at $r=2\,\rm M$  (mouth A side) while the mass flux becomes outflow at $r=-2\,\rm M$ (mouth B side). The lower panels show the magnetic flux for each model, evaluated at mouth~A ($r = 2\,\rm M$). }
\label{fig:fig1}
\end{figure*}

\begin{figure*}
  \centering
\includegraphics[width=0.8\textwidth]{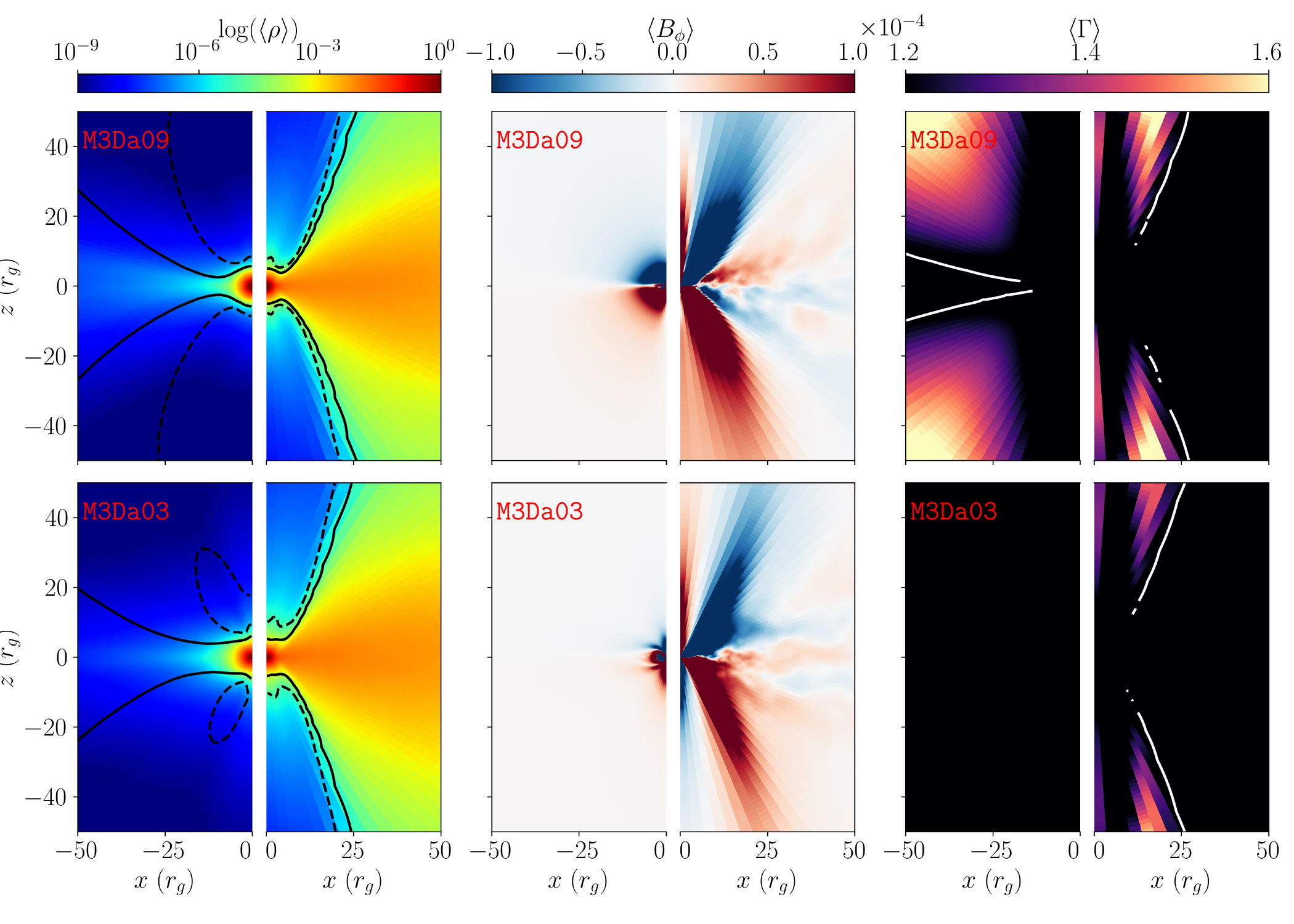}
\caption{Azimuthally and time-averaged ($t = 15{,}000$--$17{,}000\,\rm M$) logarithmic density (left), toroidal component of magnetic field, $b_\phi$ (middle), and Lorentz factor (right) for the 3D cases \texttt{M3Da03} (top) and \texttt{M3Da09} (bottom). The black contour and dashed line in the density panels correspond to $\sigma = 1$ and 10, respectively. The white contour in the Lorentz-factor panels denotes $-h u_t = 1$. For each sub-figure, the right panel ($0\,\mathrm{r_g}$ to $50\, \mathrm{r_g}$) corresponds to mouth A while the left panel ($-50\,\mathrm{r_g}$ to $0\,\mathrm{r_g}$) corresponds to mouth B.}
\label{fig:rho-b3-gamma}
\end{figure*}

\section{Result}
\subsection{Accretion flows surrounding a Kerr-like wormhole}

 In Fig.~\ref{fig:fig1}, we present the mass flux rate $\dot M$ around mouth A (upper panel) and mouth B (lower panel) for all 2D and 3D cases. Following the definitions in~\cite{2019ApJS..243...26P}, we compute the mass accretion rate evaluated at $r=\pm 2M$ and the dimensionless magnetic flux evaluated at mouth A($r= 2M$), which are defined as:
\begin{equation}
\dot M = \int_0^{2\pi} \int_0^{\pi} \rho u^r \sqrt{-g}\, d\theta\, d\phi,\\
\label{eq:accretion_flux}
\end{equation}
\begin{equation}
    \Phi_{\rm } = \frac{1}{2} \int_0^{2\pi} \int_0^{\pi} |B^r| \sqrt{-g} \, d\theta \, d\phi.
\end{equation}
We note that on the mouth A side ($r=2\,\rm M$), the mass flux is an inflow (accretion); however, on the mouth B side ($r=-2\,\rm M$), the mass flux becomes an outflow.
As shown in Figs.~\ref{fig:fig1}(a) and (b), both sides of the mass flux rate reach a quasi-steady state after a simulation time of $t \gtrsim 12,000\,\rm M$ for all cases considered. We find that the quasi-steady outflow rate around mouth B decreases with increasing values of the spin parameter $a$. For example, in the three-dimensional simulations with $a = 0.0$ and $0.9$, the time-averaged outflow rates are $\langle \dot M \rangle = 0.014$ and $0.0026$, respectively. A similar reduction in the outflow rate is also observed in the two-dimensional simulations. As shown in Fig.~\ref{fig:fig1}, the magnetic flux rate (lower panels) exhibits only minor differences among the models, with the higher-spin case showing a slightly enhanced magnetic flux. A similar result is reported by \cite{Combi:2024ehi}. All of our simulations are conducted in the standard and normal evolution (SANE) regime, with magnetic flux levels far below those required for a magnetically arrested disk (MAD) state. All of our simulations are still in the standard and normal evolution (SANE) state. The high-resolution 2D runs show the same qualitative behavior, suggesting that the matter accumulation near the throat is unlikely to be solely a numerical artifact of the lower-resolution 3D simulations.

Using the results of 3D simulations, we compare the flow structures around the models \texttt{M3Da03} and \texttt{M3Da09}. Fig.~\ref{fig:rho-b3-gamma} shows the distributions of the time-averaged density (left panels), toroidal magnetic field (middle panels), and Lorentz factor (right panels). In the left and right panels, the black solid (dashed) and white contours mark the surfaces where the magnetization reaches $\sigma \equiv b^2/\rho = 1$ (10) and where $-h u_t = 1$, respectively. Here, the negative radial domain corresponds to mouth B, while the accretion disk is initialized near mouth A in the positive radial domain. As shown in the left panel of Fig.~\ref{fig:rho-b3-gamma}, the density distributions of the two models do not exhibit significant differences, nor do the magnetized regions with $\sigma > 1$ near mouth A. Both models show a similar feature, namely that the density is higher in the central region, near the throat. In contrast, near mouth B, the \texttt{M3Da09} model exhibits a wider region with $\sigma > 1$, indicating that a more strongly magnetized outflow is developed in this case. The toroidal magnetic field in the \texttt{M3Da09} model is amplified by the frame-dragging effect induced by the rotating wormhole, as illustrated in the middle panel of Fig.~\ref{fig:rho-b3-gamma} \citep{2024JCAP...05..101J,2025EPJC...85.1178E}. As a consequence, the \texttt{M3Da09} model develops a stronger magnetic field, which is associated with a more extended high-$\sigma$ outflow region, as shown in the left panel of Fig.~\ref{fig:rho-b3-gamma}, and a modestly enhanced Lorentz factor in the right panel. Correspondingly, the gravitationally unbound region is broader in the higher-spin case.

\begin{figure}
\includegraphics[width=\linewidth]{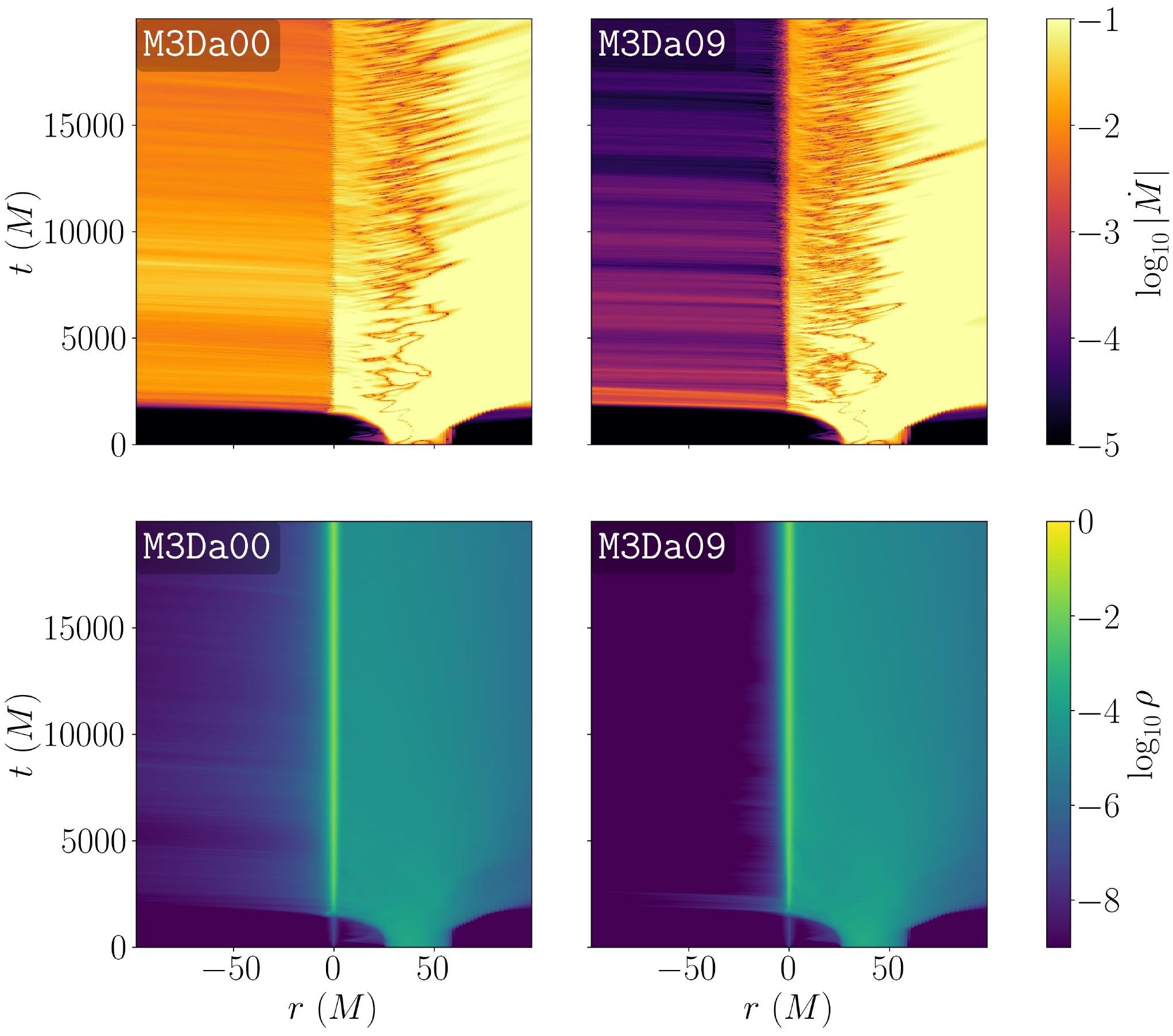}
\caption{The time evolution of the shell-averaged mass accretion (outflow) rate at mouth A (B) side $\dot M$, and density at different radii for the models \texttt{M3Da00} and \texttt{M3Da09}.}
\label{fig:spacetime}
\end{figure}

Fig.~\ref{fig:spacetime} shows the spacetime diagrams of the mass accretion  (mouth A) and outflow (mouth B) rate $\dot M$, and the density for the models \texttt{M3Da00} and \texttt{M3Da09}, where the shell-averaged quantities are defined as:
\begin{equation}
\langle f \rangle
=
\frac{\int f \, dA}{\int dA},
\qquad
dA = \sqrt{-g}\, d\theta\, d\phi .
\end{equation}

\begin{figure}
\includegraphics[width=\linewidth]{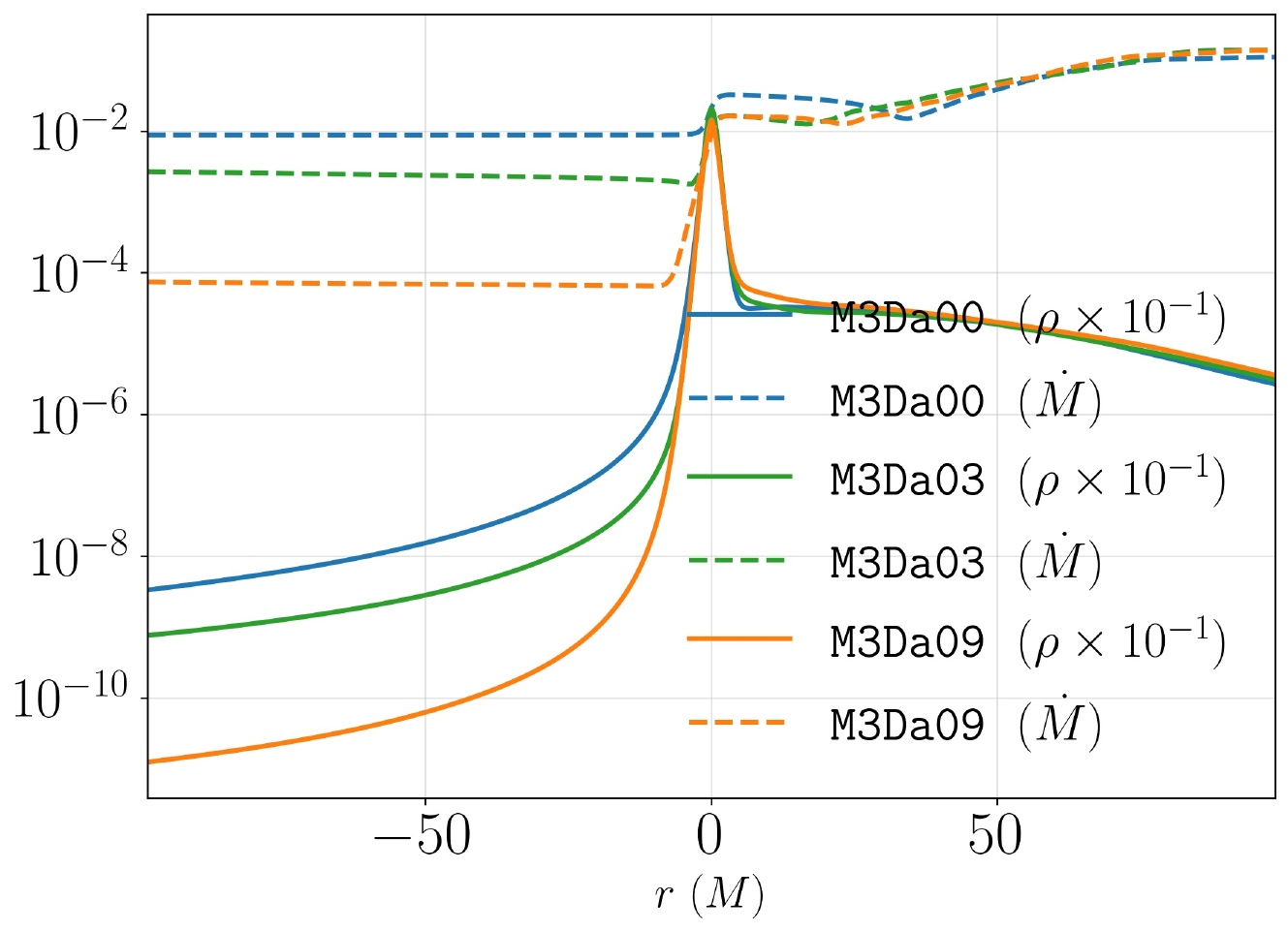}
\caption{Time-averaged radial profiles of the density and mass flux rate $\dot M$ for the models \texttt{M3Da00}, \texttt{M3Da03} and \texttt{M3Da09}. Averaged time range is from $t=15,000\,\rm M$ to $17,000\,\rm M$.}
\label{fig:rho-profile}
\end{figure}

We find that, in both models, the mass accretion rate on the mouth A side is higher than that on the mass outflow rate on the mouth B side. This mismatch between accretion and outflow rates leads to an accumulation of gas in the throat region, which can be attributed to the presence of a saddle point in the effective potential at $r = 0\,\rm M$ (see Appendix~\ref{appB} for a detailed discussion). This behavior is consistent with the centrally enhanced density discussed above, which manifests itself as the bright center band of the density diagram (lower panel) in Fig.~\ref{fig:spacetime}.

Fig.~\ref{fig:rho-profile} presents a more quantitative view of the time-averaged radial profiles for the three cases. The difference in the radially averaged mass accretion rates between mouth~A and mouth~B,
$\Delta_{\tt{Model}}=\langle \dot{M} \rangle_{\rm mouth\,A} - \langle \dot{M} \rangle_{\rm mouth\,B}$,
for the models \texttt{M3Da00} and \texttt{M3Da09} is only $\Delta_{\tt{M3Da00}}-\Delta_{\tt{M3Da09}}=3.042 \times 10^{-5}$. As a result, the peak densities at $r = 0\,\rm M$ differ only slightly between the two cases. The \texttt{M3Da03} ($a=0.3$) case shows an intermediate behavior between the
  \texttt{M3Da00} ($a=0.0$) and \texttt{M3Da09} ($a=0.9$) cases.

Although the mass outflow rate through the throat~B at larger radii appears to differ substantially between $a=0$ and $a=0.9$, we interpret this mainly as a consequence of mass conservation in response to the material accumulated near the saddle point. For larger $a$, the magnetic field on the throat-A side suppresses $\dot{M}$: near the throat, \texttt{M3Da00} is higher than \texttt{M3Da09} by $\sim 0.01$. A similar offset is present on the throat-B side, where \texttt{M3Da00} also exceeds \texttt{M3Da09} by $\sim 0.01$ (corresponding to a mass-flux difference at the $10^{-2}$--$10^{-4}$ level). This mismatch in the mass flux naturally leads to matter piling up around the saddle point. Therefore, the apparently large difference in the outflow at outer radii likely reflects the requirement to balance the mass build-up near the throat, rather than a qualitatively different outflow-driving mechanism (e.g., a systematic change in outflow speed).

In summary, our simulations show that the accretion flows in the two models share broadly similar global properties, while exhibiting systematic differences associated with the spin parameter. Although the mass accretion rate on the mouth A side consistently exceeds the mass outflow rate on the mouth B side, leading to gas accumulation near the throat due to the saddle point in the effective potential, the resulting differences in the time-averaged density profiles remain modest. The accumulation of mass near the throat that we find in the Schwarzschild-like case (\texttt{M3Da00}) is consistent with the results reported by \cite{Combi:2024ehi}. At the same time, the higher-spin case \texttt{M3Da09} develops a stronger toroidal magnetic field and a more extended high-$\sigma$ outflow region near mouth B, accompanied by a slightly enhanced Lorentz factor and a broader gravitationally unbound region. These results indicate that increasing spin primarily affects the magnetization and outflow properties, while leaving the overall accretion structure near the throat largely unchanged.

\section{GRRT image for Kerr-like wormhole}
\subsection{Shadow of Kerr-like wormhole and its critical curve}

Before presenting our results, we first briefly review the critical curve and the shadow in curved spacetime.
In Kerr-like wormhole spacetimes, the shadow is defined as the boundary on the celestial plane of a distant observer that separates photon trajectories escaping to infinity from those that remain critically bound to the central region or traverse the wormhole throat. This boundary can be characterized by unstable photon trajectories, which form the separatrix between escaping and non-escaping null geodesics \citep{2021PhRvD.103j4050K}. Owing to the separability of the Hamilton--Jacobi equation, the radial motion of photons can be described by an effective radial potential,
\begin{equation}
    \mathcal{R}(\rho)
=
\big[(\rho^2+a^2)-a\xi\big]^2
-
\Delta(\rho)\big[\eta+(\xi-a)^2\big],
\end{equation}
where $\rho=\sqrt{r^2+\ell^2}$, $\Delta(\rho)=\rho^2+a^2-2M\rho$, and $E\equiv -p_t$ and $L\equiv p_\phi$ are, respectively, the conserved energy and azimuthal angular momentum along the null geodesic, while $Q$ denotes the Carter constant \citep{1968PhRv..174.1559C}. We introduce the impact parameters $\xi\equiv L/E$ and $\eta\equiv Q/E^2$.
Unstable circular photon orbits correspond to critical solutions satisfying $\mathcal{R}(\rho_{\rm ph})=0$ and $\mathcal{R}'(\rho_{\rm ph})=0$, with instability ensured by $\mathcal{R}''(\rho_{\rm ph})>0$. Solving these conditions yields a Kerr-like branch of the critical curve when unstable photon orbits exist outside the throat, $\rho_{\rm ph}>\rho_{\rm th}=\ell$, with the corresponding impact parameters given by
\begin{equation}
\begin{aligned}
\xi(\rho_{\rm ph})
&=
\frac{
-\rho_{\rm ph}^3
+3M\rho_{\rm ph}^2
-a^2\rho_{\rm ph}
-a^2M
}
{a(\rho_{\rm ph}-M)},\\[6pt]
\eta(\rho_{\rm ph})
&=
\frac{
\rho_{\rm ph}^3
\big[
4a^2M
-\rho_{\rm ph}(\rho_{\rm ph}-3M)^2
\big]
}
{a^2(\rho_{\rm ph}-M)^2}.
\end{aligned}
\end{equation}
The projection of this one-parameter family onto the observer’s celestial plane defines a Kerr-like segment of the critical curve, which is largely insensitive to the presence of the throat. For an observer at infinity with inclination angle $\theta_0$, the celestial coordinates are given by $\alpha=-\xi/\sin\theta_0$ and $\beta=\pm\sqrt{\eta+a^2\cos^2\theta_0-\xi^2\cot^2\theta_0}$ \citep{1973blho.conf..215B}. When the throat radius exceeds the location of the innermost unstable photon orbit, the Kerr-like branch is truncated, and an additional throat branch emerges. This throat branch arises from critical photon trajectories located in the wormhole throat, $\rho=\rho_{\rm th}$. In this case, the effective potential vanishes as a consequence of the geometry of the throat, namely by condition $\Delta(\rho_{\rm th})=0$ \citep{2021PhRvD.103j4050K}. Imposing the second critical condition then requires $\mathcal{R}(\rho_{\rm th})=0$, which yields the following relation:
\begin{equation}
    \eta_{\rm th}(\xi)
=
\frac{\big[(\rho_{\rm th}^2+a^2)-a\xi\big]^2}
{\Delta(\rho_{\rm th})}
-
(\xi-a)^2.
\end{equation}

The projection of this relation onto the celestial plane defines a throat-controlled segment of the critical curve. The complete critical curve of a Kerr-like wormhole is therefore composed of a Kerr-like branch associated with unstable circular photon orbits outside the throat and a throat branch associated with photon trajectories critically limited by the wormhole geometry. The relative contribution of these branches depends sensitively on both the spin parameter and the throat size, i.e., increasing the spin enhances the asymmetry of the critical curve through frame-dragging effects, while increasing the throat radius enlarges the portion of the shadow governed by the throat branch. Although the critical curve itself is not directly observable, it represents the accumulation locus of higher-order photon trajectories and thus determines the asymptotic structure of photon rings, providing a concise and robust characterization of the shadow of Kerr-like wormholes.

\begin{figure}
  \centering
  \includegraphics[
    width=0.7\linewidth,
    trim=2 0 0 0,
    clip
  ]{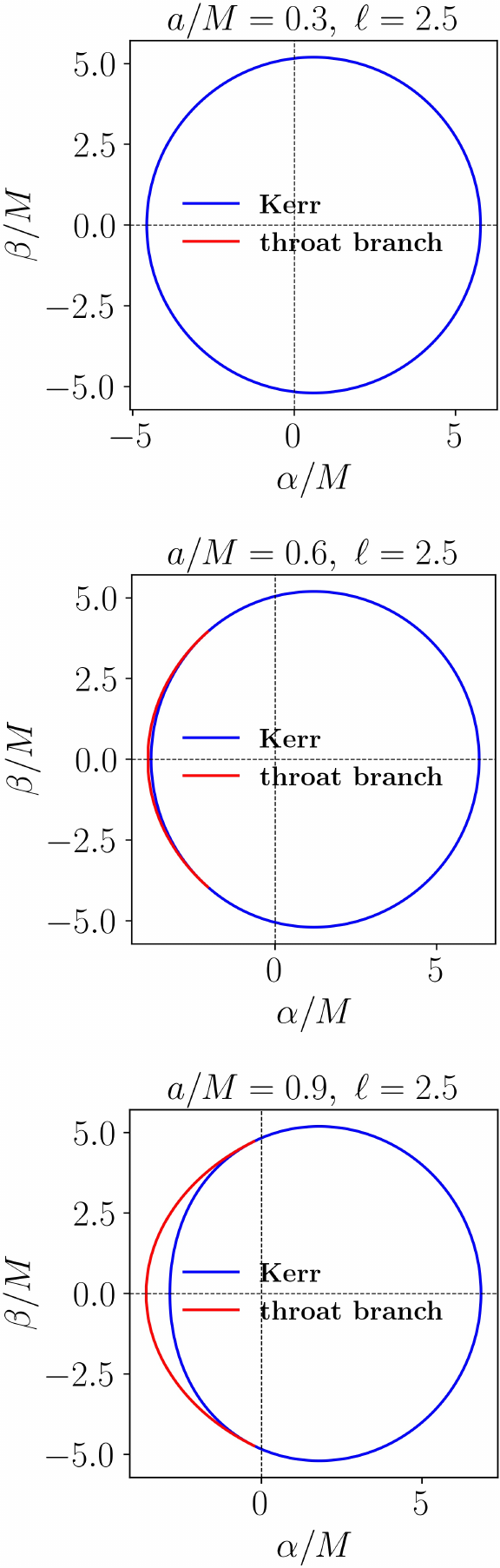}
  \caption{The figure shows the critical curves computed for $\ell = 2.5$ in different values of the spin parameter with an inclination angle of $i = 90^\circ$. Blue and red curves represent the Kerr and throat branch, respectively. }
  \label{fig:shadow}
\end{figure}

Fig.~\ref{fig:shadow} shows the analytically calculated critical curves for different values of the spin parameter, using the same throat parameter $\ell = 2.5\,\rm M$ with an inclination angle of $i = 90^\circ$.
As the spin increases, the shape of the Kerr branch is progressively modified, becoming more ``D-shaped'' asymmetry \citep{2019JCAP...03..046W,1973blho.conf..215B}. The throat branch for a fixed value of $\ell$ is also influenced by the spin, with this throat-induced effect becoming significantly more pronounced at higher spin, as shown in the lower panel of Fig.~\ref{fig:shadow}.

\subsection{GRRT image from GRMHD simulations}

\begin{figure}
\centering
\includegraphics[width=1.1\linewidth]{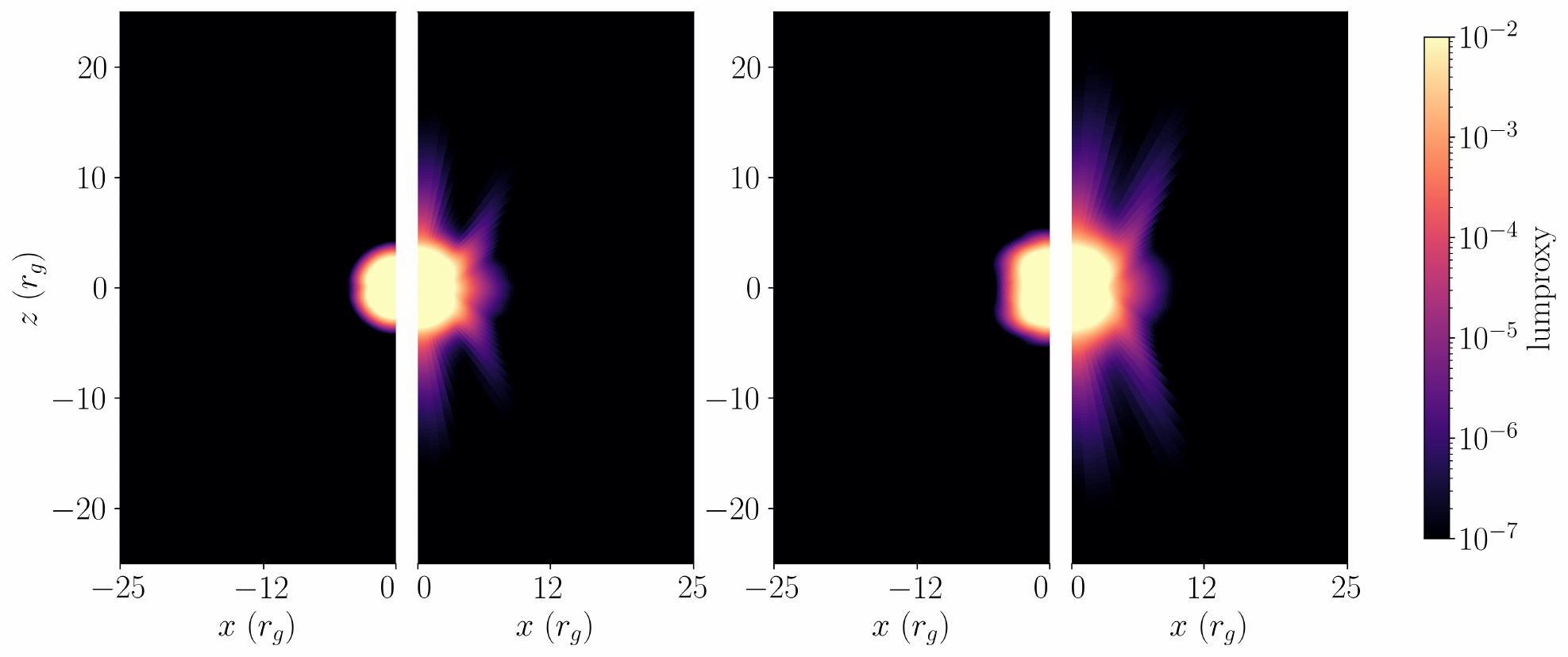}
\caption{The time and azimuthal averaged emissivity proxy $j$ for the models \texttt{M3Da03} and \texttt{M3Da09}. Averaged time range is from $15,000\,\rm M$ to $17,000\,\rm M$.}
\label{fig:emi}
\end{figure}

We first introduce a proxy for the emissivity to gain an intuitive understanding of the throat region. The proxy of emissivity $j$ is defined as \citep{2019ApJS..243...26P}:
\begin{equation}
j =  \frac{\rho^{3}}{P^{\,2}}
\exp\!\left[
-0.2
\left(
\frac{\rho^{2}}{B\,P^{\,2}}
\right)^{1/3}
\right].   
\end{equation}
As discussed above, the accumulation of mass near the mouth A leads to an increase of density in this region. The gas is also heated there, acquiring additional thermal energy through plasma processes \citep{Combi:2024ehi}. As a result of these features, the throat region exhibits higher emissivity, as shown in Fig.~\ref{fig:emi}.
This bright central emission obscures the surrounding photon ring, preventing a direct observation of the ring-like structure. We therefore use higher-order photon rings to extract the critical ring from the direct lensed emission ($n=0$) of the surrounding plasma \citep{2021A&A...646A..37V,2025arXiv251120756S}.

\begin{figure}
\centering
\includegraphics[width=1\linewidth]{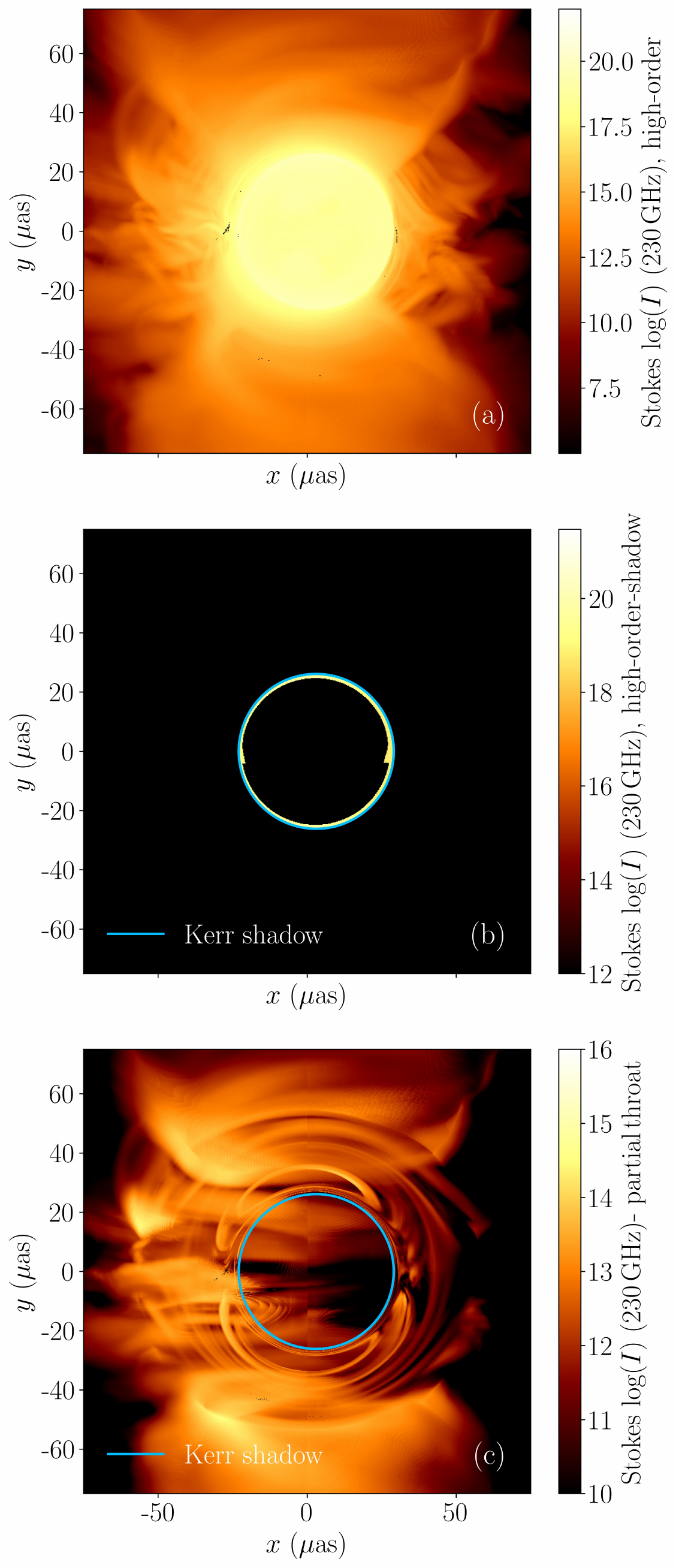}
\caption{
GRRT image of (a) the total intensity, (b) the higher-order photon rings overlaid with the critical curve, and (c) the partial throat image for the case \texttt{M3Da03} seen to mouth A (observer stayed at a far-away position in the positive radius region). For $r<2\,\rm M$, the plasma density is taken to be zero to suppress the bright central emission from the accumulated matter near the throat. The snapshot is calculated at $t=14,000\,\rm M$.  }
\label{fig:grrt03}
\end{figure}

\begin{figure}
\centering
\includegraphics[ width=1\linewidth]{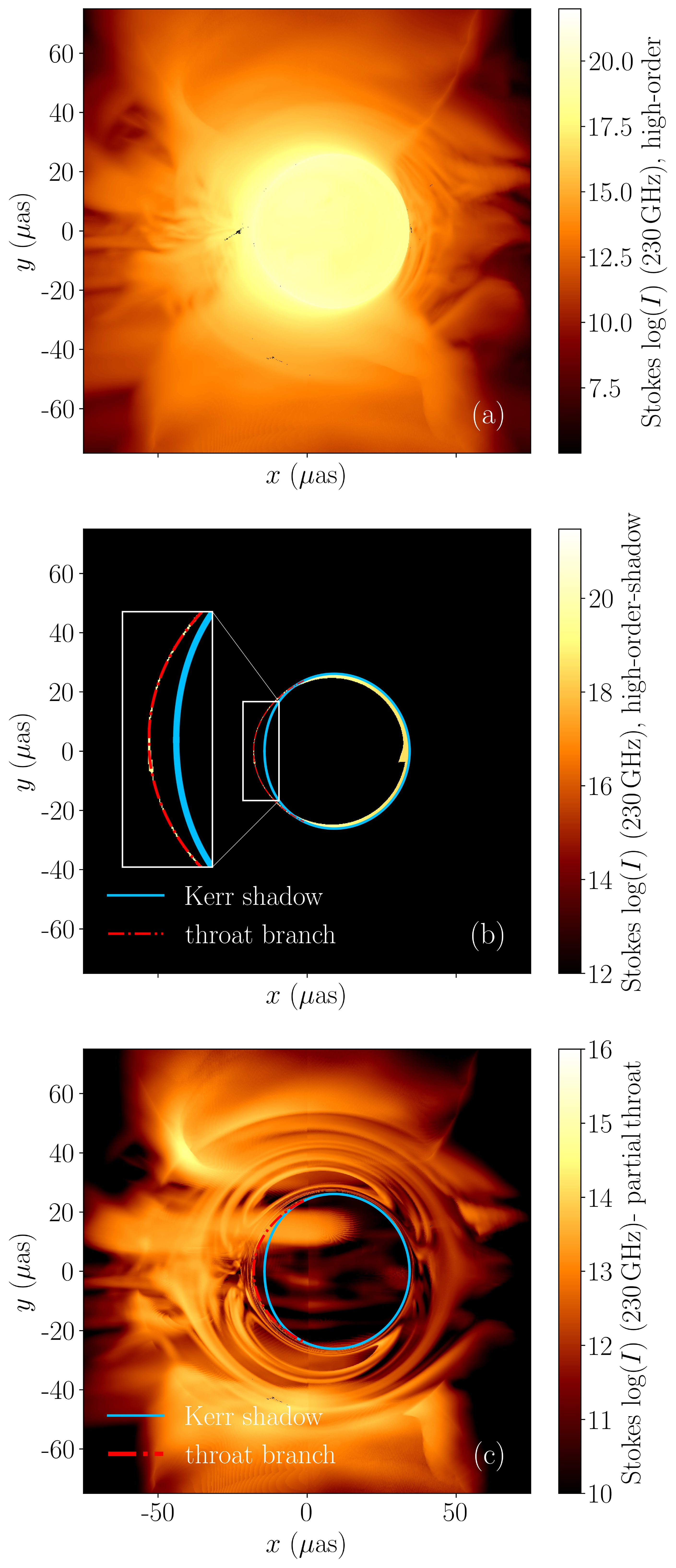}
\caption{Same as Fig.~\ref{fig:grrt03},but for higher spin case \texttt{M3Da09}. }
\label{fig:grrt09}
\end{figure}

Fig.~\ref{fig:grrt03} shows the GRRT image of the total intensity, the higher-order photon rings overlaid with the critical curve,
and the partial throat image for the case \texttt{M3Da03}.
Here, we see the mouth~A from the observer stayed at a far distance in the position of positive radius.
As shown in Fig.~\ref{fig:grrt03}, the photon-ring structure cannot be directly identified in the total image due to the bright central emission from the accumulated plasma near the throat. This shadow image is much different from that of a Kerr black hole. Therefore, we can easily distinguish between Kerr black holes and Kerr-like wormholes based on shadow images. To compare with a direct comparison with the analytically calculated critical curves shown in Fig.~\ref{fig:shadow}, we therefore extract the higher-order photon rings to mask direct lensed emission. 
For the case of \texttt{M3Da03}, no throat branch appears in the critical curve (see the top panel of Fig.~\ref{fig:shadow}). This result is confirmed by the comparison between the critical curve and the higher-order photon rings obtained from the GRRT calculation of the case of \texttt{M3Da03} presented in Fig.~\ref{fig:grrt03}. The lower panels of Fig.~\ref{fig:grrt03} and Fig.~\ref{fig:grrt09} show  GRRT images obtained with a partial cutoff of the throat region. To eliminate bright central emission coming from a plasma near a throat, we set the plasma density at zero for $r<2\,\rm M$.
In this case, the photon-ring structure becomes directly visible without further manipulation. 

Next, we examine the high-spin case \texttt{M3Da09}. In that case, the throat branch contributes to the critical curve and modifies the shape of its left-hand side. As shown in Fig.~\ref{fig:grrt09}, there is excellent agreement between the theoretical prediction (critical curve) in the throat branch and the higher-order photon rings extracted from the GRRT images. 
We note that Appendix~\ref{appenC} also presents GRRT images seen in the mouth~B, where the observer is located far away in negative radial coordinates. In this configuration, we again find bright emission around the mouth~B, accompanied by a clearly visible ring-like structure. Moreover, the images exhibit a complicated set of higher-order emission features, reminiscent of those reported for the shadow images of Kerr naked singularities \citep{2025ApJ...978...44D,2025arXiv251120756S}.
To better demonstrate these features, we also present the total intensity and the corresponding cutoff cases for different inclination angles in Appendix~\ref{appenA}. As expected, these images display a prominent ring-like structure accompanied by enhanced central brightness.

\subsection{Light curve for wormhole}

\begin{figure*}
\centering
\includegraphics[ width=1\linewidth]{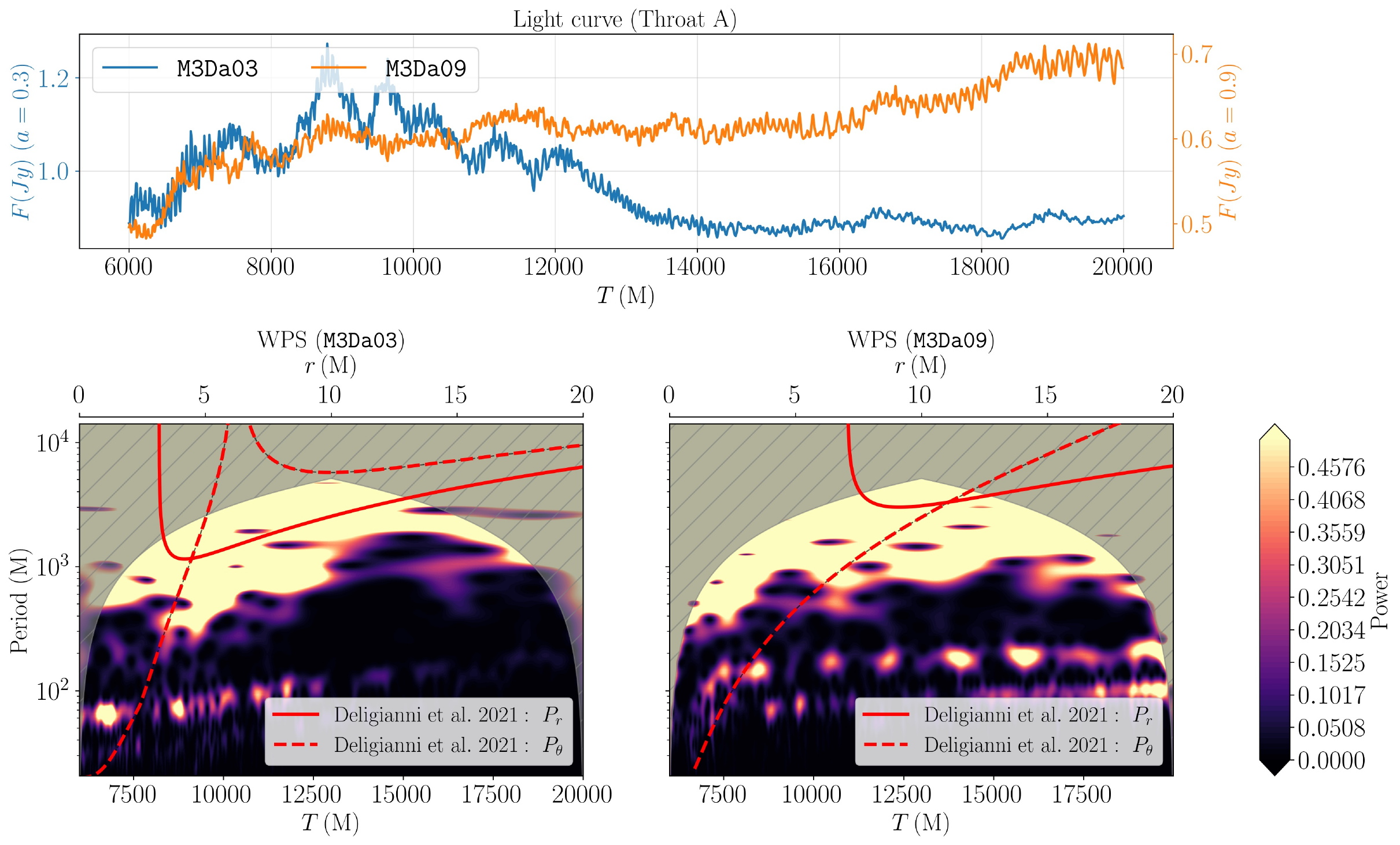}
\caption{The light curve for \texttt{M3Da03} and \texttt{M3Da09}. The lower panels show the wavelet power spectrum over the same time interval as in the upper panel, together with the analytical period--radius relation ($P_r(r)$ and $P_\theta(r)$) from \citet{2021PhRvD.104b4048D}. We find that the dominant oscillation period is associated with the inner
  region of the flow, i.e., the vicinity of the throat, with the theoretical
  curve ($P_\theta(r)$) overlapping the bright wavelet-power band near ($r \sim 3\,\rm M$). }
\label{fig:lightcurve}
\end{figure*}

From our GRRT calculations, we directly obtain the light curves for the Kerr-like wormhole. Fig.~\ref{fig:lightcurve} shows the light curves for the cases of \texttt{M3Da03} and \texttt{M3Da09} in the upper panel, while the lower panels display the corresponding wavelet power spectra.
We find a characteristic quasi-periodic oscillation (QPO) in the light curves, with a corresponding typical period ($\sim 100\,\rm M$) clearly identified in the wavelet power spectra.

The above QPO signature can be naturally interpreted within the epicyclic resonance framework for accretion flows in rotating traversable wormhole spacetimes \citep{2021PhRvD.104b4048D}. 
In this picture, the nearly circular motion of plasma elements admits small-to-moderate perturbations that can be decomposed into radial and vertical epicyclic modes with eigenfrequencies $\omega_r$ and $\omega_\theta$ (see equations (17) and (18) in \cite{2021PhRvD.104b4048D}), whose ordering and possible resonant couplings may be substantially richer than in the Kerr black hole case. In particular, the wormhole geometry can give rise to a non-trivial effective potential structure in the vicinity of the throat, allowing for (meta-)stable bounded motion. Hence, the throat-adjacent potential structure provides a natural characteristic timescale for quasi-
  periodic motion. Whether this motion produces a long-lived observable modulation depends on the continued excitation and emissivity contrast of the gas near the throat. In our case, the dominant high-frequency QPO feature is better aligned with the vertical epicyclic timescale (dashed lines), which corresponds to the high-frequency branch of the $\omega_\theta$ profile evaluated in the inner region near the throat. It suggests that the observed modulation is closely associated with the $\omega_\theta$ mode (or combinations associated with it). 
Our GRRT results provide a concrete radiative realization of this dynamics. The potential cavity acts as a temporary reservoir in which matter can undergo quasi-periodic radial oscillations, which may be viewed as the non-linear, finite-amplitude extension of the radial epicyclic motion \citep{2021PhRvD.104b4048D}. In an idealized test-particle limit without dissipation, such trajectories remain trapped and oscillatory. In a realistic magnetized plasma, however, interactions with the turbulent inflow and magnetic stresses can lead to intermittent energy exchange, gradually enabling portions of the oscillating material to cross the potential barrier and escape.

Crucially, when trapped matter swings back toward the inflow, it can episodically compress, shock, or enhance dissipation at the interface with the hot, magnetized accretion stream, producing transient emissivity boosts. The superposition of these recurrent brightening episodes imprints a quasi-periodic modulation onto the light curve, which appears as the coherent timescale identified in the wavelet power spectra. In this sense, the observed QPO can be interpreted as the observable (radiative) manifestation of non-linear epicyclic dynamics enabled by the throat-adjacent potential structure unique to wormhole spacetimes.

\section{Summary and Discussion}

In this work, we performed a series of 2D and 3D GRMHD simulations together with GRRT calculations to investigate the accretion dynamics, horizon-scale image structure, and time-dependent radiative signatures of Kerr-like wormholes. Although the overall accretion morphology closely resembles that of Kerr black holes at leading order, the presence of a wormhole throat introduces distinctive features in the accumulation of matter in the accretion flow dynamics, which reflects on the horizon-scale images and temporal variability of total intensity. In the following, we summarize our findings and discuss their potential observational implications.

\begin{enumerate}

    \item 
    Our simulations reveal a general tendency for matter to accumulate near the wormhole throat. The mass flux rate on the inflow side of mouth~A systematically exceeds that on the outflow side of mouth~B, leading to gas build-up in the throat region. This behavior can be naturally understood as a consequence of the saddle point in the effective potential at the throat, which traps matter and enhances the central density. Despite this accumulation, the time-averaged density profiles remain broadly similar across different spin values, indicating that the large-scale accretion structure near the throat is relatively insensitive to spin.

    \item 
    The spin parameter primarily affects the magnetization and outflow properties rather than the density distribution. As the spin increases, the frame-dragging effect amplifies the toroidal magnetic field and produces a more extended, highly magnetized outflow region near mouth~B. This is accompanied by a modest increase in the Lorentz factor and a broader gravitationally unbound region, consistent with enhanced magnetic winding and more efficient acceleration in rapidly rotating spacetime.

    \item 
    Analytically, the critical curve of a Kerr-like wormhole consists of a Kerr-like branch associated with unstable circular photon orbits outside the throat and, when the throat radius is sufficiently large, an additional throat branch arising from photon trajectories critically constrained by the throat geometry. Our GRRT results confirm this picture. In the moderate-spin case (\texttt{M3Da03}), the critical curve is entirely governed by the Kerr-like branch, whereas in the high-spin case (\texttt{M3Da09}), the throat branch significantly modifies the critical curve in an asymmetric structure. The close agreement between the analytical critical curves and the higher-order photon rings extracted from GRRT images demonstrates that higher-order photon rings provide a robust probe of the underlying photon dynamics, even when the total image is dominated by bright central emission.

    \item 
    Matter accumulation and plasma heating near the throat lead to a strong central emissivity component, which can obscure the photon-ring structure in total intensity images. This horizon-scale image is much different from that of a Kerr black hole.
    By isolating higher-order photon rings or imposing a partial throat cutoff in the GRRT calculation, the ring structure becomes directly visible and can be reliably compared with the analytical critical curve. This highlights the importance of separating different orders of photon ring contributions when interpreting images of horizonless compact objects.

    \item 
    The GRRT light curves show QPO-like variability during the analyzed time interval. We attribute the
  characteristic timescale of this variability to the presence of a potential cavity near the wormhole
  throat, where accumulated gas can undergo quasi-periodic motion. However, the observable modulation in
  the light curve depends not only on this geometric timescale, but also on the continued excitation and
  emissivity contrast of the gas near the throat. In a realistic magnetized plasma, interactions with the
  turbulent inflow, shocks, and magnetic stresses can intermittently enhance the emissivity and produce
  transient brightenings.
  Thus, the throat-adjacent potential structure provides a natural mechanism for setting the QPO-like
  timescale, while the amplitude and persistence of the signal remain flow-dependent. Therefore, our
  results should be interpreted as evidence for QPO-like variability over the simulated interval, rather
  than as a demonstration of a persistent long-term QPO.

\end{enumerate}

In summary, Kerr-like wormholes can closely mimic Kerr black holes in their overall accretion structure, although leaving distinct imprints in horizon-scale image morphology, photon ring structure, and time-domain variability. The emergence of a throat branch in the critical curve, the behavior of higher-order photon rings, and the presence of QPO-like signatures provide complementary diagnostics that help to distinguish wormholes from black holes. The matter accumulation near the throat may form an optically thick, effectively thermalizing layer with blackbody-like surface emission, which would be testable and constrained by coordinated multi-wavelength data and infrared limits\citep[e.g.,][]{2022ApJ...930L..17E,2024A&A...689A.197D}.

We note that, as shown by \citep{Combi:2024ehi}, wind launching on mouth B depends on the throat size. It is therefore possible that a different throat parameter could lead to wind launching even in the lower-spin case ($a=0.3$). Similarly, accretion flow properties in our simulations show the SANE state only. A more magnetically dominated regime, MAD state, may modify the wind-launching behavior by changing the magnetic-flux accumulation and field strength near the throat. A systematic exploration of these effects would be crucial for future studies. From an observational point of view, the polarization would be an important observable that we have not investigated yet. In future work, polarized radiative transfer and direct comparisons with EHT observations, will be crucial for assessing the observational viability of these signatures.

\begin{acknowledgements}
We thank Xufan Hu, Akhil Uniyal, and Indu K. Dihingia for useful discussions.
This research is supported by the National Key R\&D Program of China (grant No. 2023YFE0101200), the National Natural Science Foundation of China (grant Nos. 12273022, 12511540053), and the Shanghai Municipality orientation program of Basic Research for International Scientists (grant No. 22JC1410600).
The simulations were analyzed on the TDLI-Astro cluster at Shanghai Jiao Tong University.
\end{acknowledgements}

\begin{appendix}

\section{GRRT image for different inclination angles}\label{appenA}
\renewcommand{\thefigure}{A1}

\begin{figure}
    \centering
    \includegraphics[width=\linewidth]{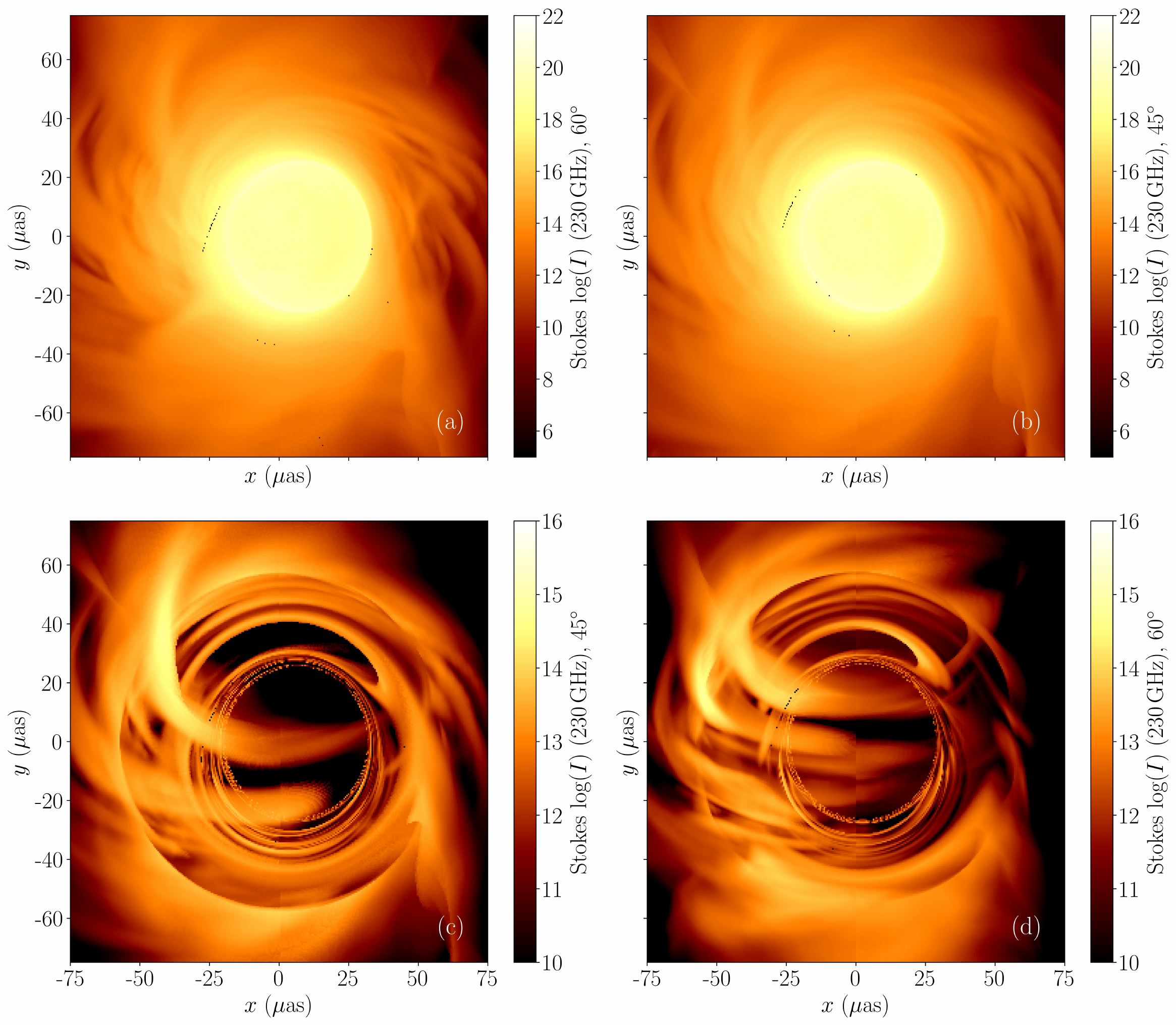}
\caption{GRRT images at $t=14{,}000\,\rm M$ for \texttt{M3Da09} at inclination angles of $45^\circ$ and $60^\circ$. The upper panels (a) and (b) show the total intensity, while the lower panels (c) and (d) display the corresponding partial throat images(same as Fig.~\ref{fig:grrt09}(c) but for different angles).
}
    \label{fig:grrt-angles}
\end{figure}

Here we present the results for different inclination angles of the high-spin model \texttt{M3Da09}. As shown in Fig.~\ref{fig:grrt-angles}, panels (c) and (d) reveal a complex set of fine sub-ring features. These detailed sub-ring structures arise from the underlying spacetime geometry shaped by the wormhole throat\citep{2021A&A...646A..37V}.

\section{Effective-Potential Construction for Timelike Geodesics}
\label{appB}

\renewcommand{\thefigure}{B1}
In this part, we describe the construction of the effective-potential-like curves used in this work to characterize radial motion of timelike geodesics in the Kerr-like wormhole spacetime. This construction provides a convenient way to identify radial turning points and bound orbital regions, and enables a direct comparison of orbital properties across different values of the spin parameter.

We consider equatorial ($\theta=\pi/2$) timelike geodesics in a Kerr-like wormhole geometry, where the Boyer-Lindquist radial coordinate is regularized as
\begin{equation}
\rho \equiv \sqrt{r^2+\ell^2},
\end{equation}
with $\ell$ denoting the wormhole throat parameter. The radial equation of motion can be written in the separable form
\begin{equation}
\Sigma^2\left(\frac{d\rho}{d\lambda}\right)^2 = R(\rho),
\end{equation}
where $\lambda$ is an affine parameter and the radial potential $R(\rho)$ is given by
\begin{equation}
R(\rho)=\left[E(\rho^2+a^2)-aL\right]^2
-\Delta(\rho)\left(\mu^2\rho^2+(L-aE)^2\right),
\label{eq:R_rho}
\end{equation}
with
\begin{equation}
\Delta(\rho)=\rho^2-2M\rho+a^2.
\end{equation}
Here $E$ and $L$ are the conserved energy and azimuthal angular momentum per unit mass, $\mu$ denotes the particle mass parameter ($\mu=1$ for timelike geodesics), and $a$ is the spin parameter. Throughout this section, we set $M=1$.

For a fixed angular momentum $L$, radial turning points are determined by the condition
\begin{equation}
R(\rho)=0.
\end{equation}
At fixed $\rho$, Eq.~\eqref{eq:R_rho} is a quadratic equation in the conserved energy $E$,
\begin{equation}
A(\rho)E^2+B(\rho)E+C(\rho)=0,
\end{equation}
where the coefficients depend only on $\rho$, $a$, $L$, and $\mu$. This equation admits two solution branches,
\begin{equation}
E_{\pm}(\rho)=\frac{-B(\rho)\pm\sqrt{B(\rho)^2-4A(\rho)C(\rho)}}{2A(\rho)},
\end{equation}
provided that the discriminant is non-negative. The existence of real solutions, therefore, determines the radial region accessible to a particle with given conserved quantities.

\begin{figure}
    \centering
    \includegraphics[width=0.7\linewidth]{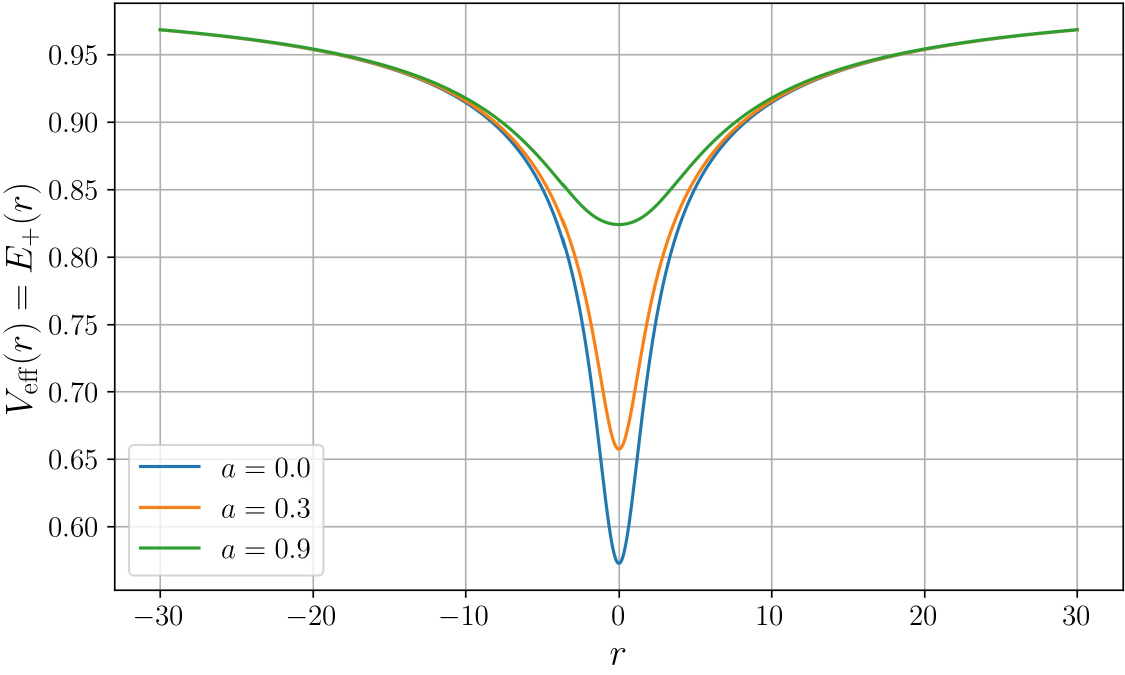}
\caption{ The effective potential for models with different spin parameters at fixed $\ell = 2.5$.
}
    \label{fig:potential}
\end{figure}

In this work, we define an effective-potential-like function by identifying the upper branch $E_{+}(\rho)$ as: 
\begin{equation}
V_{\rm eff}(r)\equiv E_{+}\!\left(\rho=\sqrt{r^2+\ell^2}\right).
\end{equation}
Although $V_{\rm eff}$ shown in Fig.~\ref{fig:potential} is not a potential in the strict Newtonian sense, it plays an analogous role: for a particle with fixed angular momentum $L$ and energy $E$, radial motion is allowed only in regions where $E\ge V_{\rm eff}(r)$, while radial turning points correspond to intersections between $E$ and $V_{\rm eff}(r)$.

\section{GRRT image from Mouth B}\label{appenC}
\renewcommand{\thefigure}{C1}

\begin{figure}
    \centering
    \includegraphics[width=\linewidth]{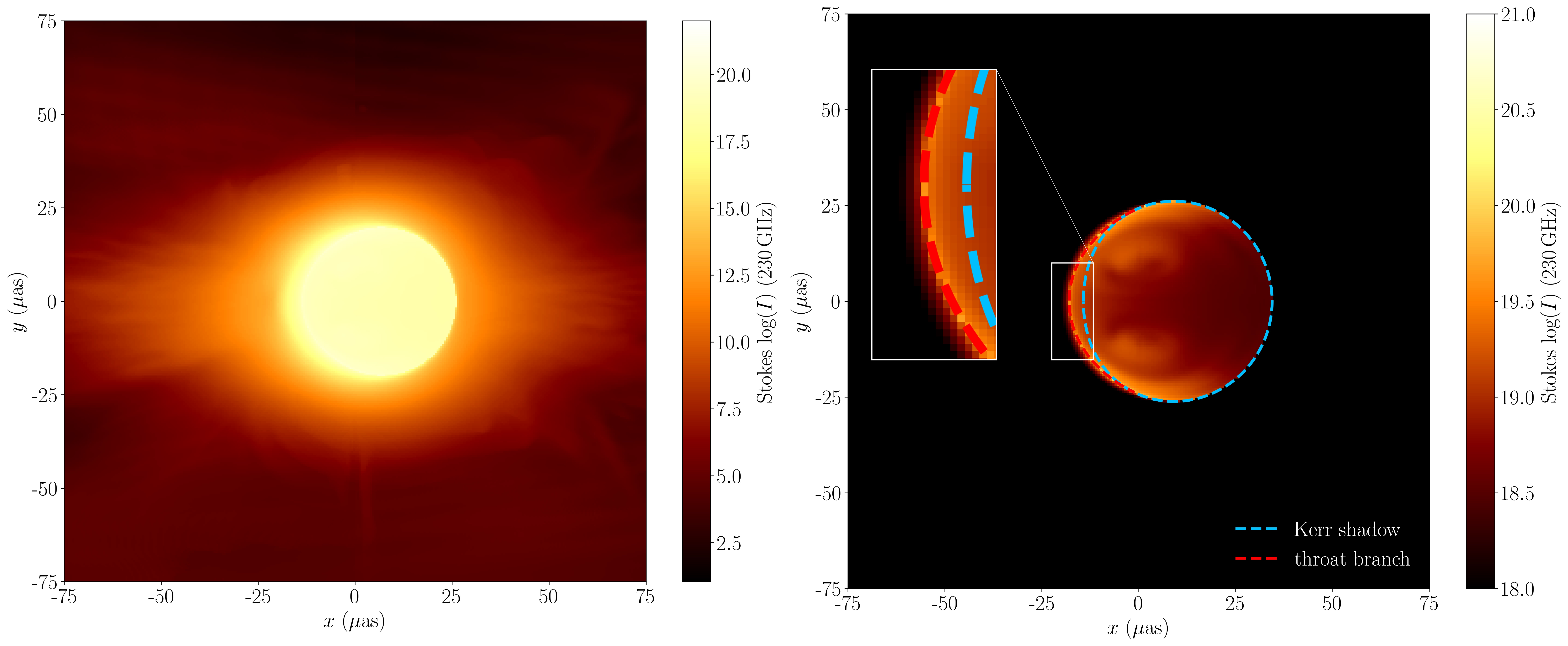}
\caption{The time-averaged GRRT images observed from the mouth B side are shown together with the overlaid critical curve for \texttt{M3Da09}. We note that the different ranges are used for the color bars. Averaged time range is $[15,000\,\rm M-17,000\,\rm M]$.
}
    \label{fig:grrt-throat B}
\end{figure}

By placing the observer on the mouth B side, we can directly obtain the corresponding GRRT images from this viewpoint, as shown in Fig.~\ref{fig:grrt-throat B}. We find that there are no strong jet formations on this side; instead, the emission is more extended along the radial direction, corresponding to an outflow-dominated structure.

\section{Quality factor}\label{appenD}
\renewcommand{\thefigure}{D1}
Here, we investigate whether our simulations can capture the MRI in the simulation well.
In practice, the fastest-growing MRI mode is considered well resolved when the MRI quality factor $Q \gtrsim 6 $ \citep{2004ApJ...605..321S}. 
Figure~\ref{fig:quality factor} shows the MRI quality factor calculated in all three
 directions for model \texttt{M3Da09} at $t=17{,}000\,\rm M$, for
  comparison with \cite{Combi:2024ehi}. It is indicated that the MRI is reasonably resolved (e.g., Figure~9 in \cite{Combi:2024ehi}) throughout most of the domain. Therefore, our grid resolution is enough to capture the evolution of simulations properly.

\begin{figure}
    \centering
    \includegraphics[width=0.8\linewidth]{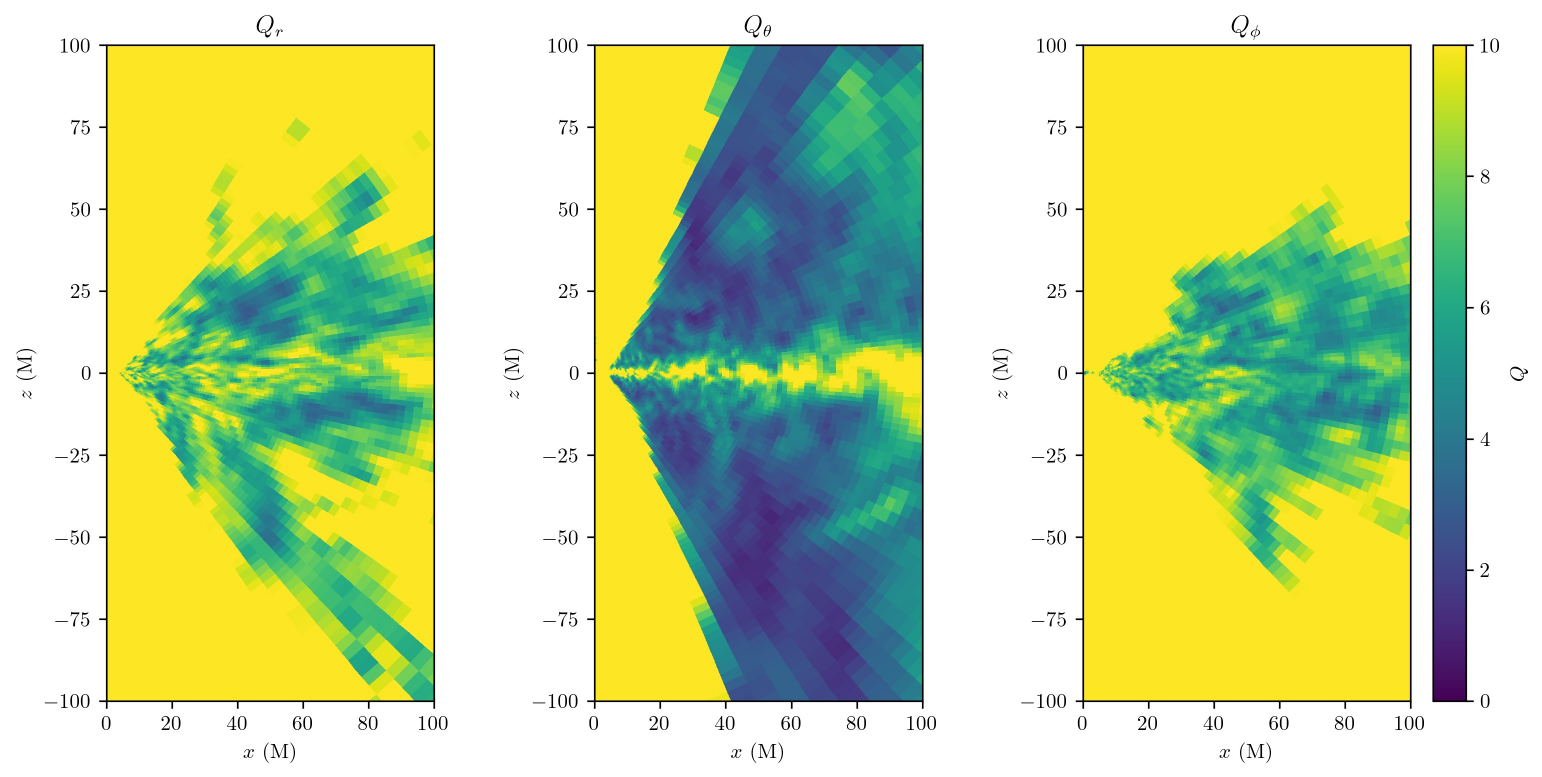}
\caption{MRI quality factor for model $\tt{M3Da09}$ at $t = 17{,}000\,\rm M$. 
}
    \label{fig:quality factor}
\end{figure}

\end{appendix}

\bibliography{sample701}{}
\bibliographystyle{aasjournalv7}



\end{document}